\documentclass{ws-ijmpa}
\usepackage{hyperref}
\usepackage[super,compress]{cite}
\usepackage{graphicx}

\usepackage{bbm}
\usepackage{multirow}

\DeclareMathSymbol{\mg}{\mathrel}{symbols}{"1D}

\newcommand{\bes}{\begin{split}}
\newcommand{\ees}{\end{split}}

\newcommand{\gd}{\delta}
\renewcommand{\ge}{\epsilon}
\newcommand{\gve}{\varepsilon}

\newcommand{\gk}{\kappa}

\newcommand{\gt}{\tau}
\newcommand{\go}{\omega}

\newcommand{\gG}{\Gamma}

\newcommand{\gF}{\Phi}

\newcommand{\gX}{\Xi}
\newcommand{\gL}{\Lambda}

\newcommand{\gPs}{\Psi}

\newcommand{\gY}{\Upsilon}
\newcommand{\cA}{{\cal A}}

\newcommand{\ra}{\rightarrow}

\newcommand{\beq}{\begin{equation}}
\newcommand{\eeq}{\end{equation}}
\newcommand{\barr}{\begin{array}}
\newcommand{\earr}{\end{array}}
\newcommand{\equ}[1]{\begin{gather} #1 \end{gather}}

\newcommand{\equa}[1]{\begin{align} #1 \end{align}}

\newcommand{\arry}[2]{\begin{array}{#1} #2 \end{array}}

\newcommand{\non}{\nonumber}
\newcommand{\sfrac}[2]{\mbox{$\frac{#1}{#2}$}}
\newcounter{oldcounter}

\newcommand{\bgo}{{\bar\omega}}

\newcommand{\tga}{{\tilde \alpha}}

\newcommand{\Intr}{\mathbbm{Z}}

\newcommand{\ba}[2]{\[\begin{array}{#2}\label{#1}}
\newcommand{\ea}{\end{array}\]}
\newcommand{\be}{\begin{equation}}
\newcommand{\ee}{\end{equation}}
\newcommand{\bea}{\begin{eqnarray}}
\newcommand{\eea}{\end{eqnarray}}

\newcommand{\sm}{{\,\mbox{-}}}

\renewcommand{\draftnote}{}

\begin{document}
\markboth{S.~Groot Nibbelink}{GLSM resolutions of torsional heterotic 
 ${\Intr_2\times\Intr_2}$ orbifolds}

%
\catchline{}{}{}{}{}
%

\title{GLSM resolutions of torsional heterotic $\boldsymbol{\Intr_2\times\Intr_2}$ orbifolds}

\author{Stefan Groot Nibbelink}
\address{School of Engineering and Applied Sciences, Rotterdam University of Applied Sciences, \\ 
G.J.\ de Jonghweg 4 - 6, 3015 GG Rotterdam, the Netherlands
 \\[1ex]
Research Centre Innovations in Care, Rotterdam University of Applied Sciences, \\ 
Postbus 25035, 3001 HA Rotterdam, the Netherlands\\
s.groot.nibbelink@hr.nl}

\maketitle

\begin{history}
14 November 2023
\end{history}

\begin{abstract}

Heterotic toriodal $\Intr_2\times\Intr_2$ orbifolds may possess discrete torsion between the two defining orbifold twists in the form of additional cocycle factors in their one--loop partition functions. 
Using Gauged Linear Sigma Models (GLSMs) the consequences of discrete torsion can be uncovered when the orbifold is smoothed out by switching on appropriate blowup modes. 
Here blowup modes with twisted oscillator excitations are chosen to reproduce bundles that are close to the standard embedding without torsion. 
The orbifold resolutions with discrete torsion are distinguished from resolutions without torsion, since they require NS5--branes at their exceptional cycles. 

\end{abstract}

\ccode{PACS numbers: 11.15.-q, 11.25.-w, 11.30.Pb, 12.10.-g}

\section{Introduction}
\label{sc:Introduction}

String theory may provide the ultimate theory of nature unifying both gravitational and gauge interactions among chiral fermions.
Heterotic string compactifications using $\Intr_2\times \Intr_2$ orbifolds of six--dimensional tori $T^6$ are among the most studied string constructions to date with appealing phenomenological properties like three generations of quarks and leptons with the potential to provide realistic Yukawa interactions. 
Such orbifold models have been studied both using the free fermionic formulation, see {\em e.g.}\ refs.\cite{Faraggi:1989ka,Faraggi:1991jr,Faraggi:1992fa, Cleaver1999,Faraggi:2006qa,Faraggi:2017cnh}\,, or using bosonic constructions~\cite{Lebedev:2006kn,Lebedev:2007hv,Lebedev:2008un,Blaszczyk:2009in}
 as well as their smooth resolutions\cite{Blaszczyk:2010db}\,.

One of the theoretically appealing features of toroidal $\Intr_2\times\Intr_2$ orbifolds is that they admit an exact conformal field theory description. 
In particular, their one loop partition function can be written down explicitly as a sum over untwisted and twisted orbifold sectors. 
A well--known feature of $\Intr_2\times\Intr_2$ orbifolds is that they may possess discrete torsion\cite{Vafa:1986wx,Vafa:1994}\,. 
In the model with discrete torsion precisely the conjugate twisted states survive the orbifold projections as compared to the model without torsion.

This proceedings investigates what happens to the discrete torsion between orbifold twists if one fully resolves the orbifolds. 
Gauged Linear Sigma Models (GLSMs) are used to do so, as they are a convenient framework interpolating between the singular orbifold limit and smooth compactifications. 
This work is therefore closely related to ref.~\cite{Faraggi:2022gkt} by the author. 
(See also the thesis \cite{Hurtado-Heredia:2023lho}\,.) 
The main difference is that in that reference non--oscillator blowup modes were considered, while in this proceedings twisted states with oscillators are used as the blowup modes. 
Apart from investigating how general the results of that reference are, this investigation is particularly interesting as standard embedding resolutions make use of oscillator blowup modes.

This might seem like a minor difference, but is in fact very important. 
The resolution of the $T^6/\Intr_2\times\Intr_2$ without discrete torsion described in this proceedings corresponds to the standard embedding in the heterotic literature and allows for a $(2,2)$ supersymmetric description of the interacting part of the worldsheet theory. 
However, as is shown in this proceedings the resolution GLSM with discrete torsion in the heterotic context does not possess $(2,2)$ supersymmetry as certain charge assignments of the fermi superfields needs to be opposite to that of the chiral multiplets. 
That this has to be the case, is dictated by the fact that oppositely charged twisted states exist in the orbifold model with torsion. Indeed, as we will see it are these twisted states which   can be used as blowup modes.
Thus even in the most basis case, associated to the standard embedding, the resolved hetorotic orbifold model with torsion requires the introduction of non--perturbative physics encoded in field dependent FI--terms, as was found for the resolution GLSM with non--oscillator blowup modes\cite{Faraggi:2022gkt}\,.

The proceedings is structured as follows: 
Section~\ref{sc:Z22orbifolds} gives a brief introduction to $\Intr_2\times\Intr_2$ orbifolds without and with discrete torsion and labels the twisted states with oscillators. 
Section~\ref{sc:T6Z22glsms} describes resolutions of the compact $T^6/\Intr_2\times\Intr_2$ using a $(0,2)$ GLSM formalism starting from a triple elliptic curve to parameterise the six--torus. 
Section~\ref{sc:T6Z22torsion} gives the GLSM description of the resolved orbifold with torsion. 
The proceedings is completed with a discussion section comparing the results obtained here with existing literature.

\section{Discrete torsion on $\boldsymbol{\Intr_2\times\Intr_2}$ orbifolds}
\label{sc:Z22orbifolds}

This section describes some essential properties of heterotic $\Intr_2\times\Intr_2$ orbifolds necessary  to investigate their resolutions using GLSM techniques. 
A more complete review of heterotic orbifolds can be found, {\em e.g.}, in refs.\cite{Dixon:1985jw,Dixon:1986jc,ibanez_88,Ibanez:1987sn,Nilles:2011aj} in general and in,  {\em e.g.}, refs.\cite{Forste:2004ie,Faraggi:2004rq,Donagi2004,Donagi2008,Blaszczyk:2009in} for $\Intr_2\times\Intr_2$ orbifolds in particular.

The standard $T^6/\Intr_2\times\Intr_2$ orbifold is defined by a six--dimensional torus $T^6$ with complex coordinates $z=(z_u)$, where $u=1,2,3$ labels the three two--torus directions, on which two twists act as 
\equ{
z \ra e^{2\pi i\,v_g} \, z~, 
\qquad 
v_{g}= t_{1}\, v_{1}+t_{2}\, v_{2}~,
}
where the twist vectors are taken to be
\equ{
v_{1}=\big(0, \sfrac{1}{2},\sm\sfrac{1}{2}\big)~,
\qquad 
v_{2}=\big(\sm\sfrac{1}{2},0, \sfrac{1}{2}\big)~. 
}
The labels, $t_1, t_2 = 0,1,$ distinguish the untwisted ($t_1=t_2=0$) and three twisted sectors. These orbifold twists are extended to act as shifts on the gauge lattice via the gauge shift embedding
\equ{
V_{g}=t_{1}\, V_{1}+t_{2}\, V_{2}~. 
}
The orbifold standard embedding takes the gauge shift vectors identical to the twist vectors augmented with the appropriate number of zero entries so that their total number of entries equals 16: 
\equ{ \label{eq:SpecificGaugeShifts} 
V_{1}=\big(0, \sfrac{1}{2},\sm\sfrac{1}{2}, 0^{5}\big)\big(0^{8}\big)~,
\qquad 
V_{2}=\big(\sm\sfrac{1}{2},0, \sfrac{1}{2}, 0^{5}\big)\big(0^{8}\big)~
}
embedded in the $E_8\times E_8$ root lattice.

The full one loop partition function is formed out of modular covariant blocks. These blocks are labeled by constructing elements $g, h$ of the orbifold group, defining the distinct sectors, and the projecting elements $g', h'$, implementing orbifold projections. These elements are summed to form the full modular invariant partition function. By this construction the orbifold partition function is uniquely defined up to certain discrete torsion phases\cite{Vafa:1986wx,Vafa:1994}\,.
In particular, the discrete torsion between the orbifold twists reads  
\equ{
\gF^{\!\times}{}^{t_1,t_2}_{t_1',t_2'} = e^{\pi i\, \gve^{\!\times} (t_{1}^{\phantom{\prime}} t_{2}^{\prime}-t_{2}^{\phantom{\prime}} t_{1}^{\prime})}~. 
}
Discrete torsion is absent for $\gve^{\!\times}=0$ and present for $\gve^{\!\times}=1$. 

Instead of introducing this non--trivial discrete torsion phase in  the full partition function, one can equivalently consider a so--called brother model~\cite{Ploger:2007iq} with gauge shift vectors 
\equ{ \label{eq:BrotherGaugeShifts} 
V^{\!\times}_{1}= -V_1=\big(0, \sm\sfrac{1}{2},\sfrac{1}{2}, 0^{5}\big)\big(0^{8}\big)~,
\qquad 
V^{\!\times}_{2}=-V_2=\big(\sfrac{1}{2},0, \sm\sfrac{1}{2}, 0^{5}\big)\big(0^{8}\big)~, 
}
that are opposite to~\eqref{eq:SpecificGaugeShifts} and therefore differ by $E_8\times E_8$ lattice vectors. 

The shifted momenta 
\equ{ \label{eq:ShiftedMomenta} 
p_{g}=p+v_{g}~, 
\qquad 
P_{g}=P+V_{g}~, 
}
identify the states in the orbifold spectrum up to possible twisted left--moving oscillator excitations 
$\tga{}^u_{-1/2}$ and $\overline{\tga}{}^u_{-1/2}$.  
Here $p \in V_4\oplus S_4$ and $P \in(O_8\oplus S_8)\otimes (O_8\oplus S_8)$, where $V_n$, $O_n$ and $S_n$ denote the vector, adjoint and spinor lattices in $n$ dimensions, respectively.

The shifted momenta of massless states $|p_g, P_g\rangle$ with possible oscilators in the physical spectrum are characterised by the requirements 
\equ{ 
\frac{1}{2}\,p_{g}^{2} = \frac{1}{2}-\delta c_g~, 
\qquad 
\frac{1}{2}\, P_{g}^{2} = 1-\delta c_g- \go_{g} \cdot \widetilde{N}_{g}-\bgo_{g} \cdot \overline{\widetilde{N}}_{g}~,  
}
where the orbifold vacuum shift 
\equ{
\delta c_g = \frac{1}{2} \sum_u \go_{g,u}(1-\go_{g,u})
}
is defined in terms of $\go_{g,u} \equiv (v_g)_u$ and $\bgo_{g,u} \equiv - (v_g)_u$ which satisfy the inequalities: $0 < \go_{g,u}, \bgo_{g,u} \leq 1$. Finally,  the number operators $(\widetilde{N}_g)_u$ and $(\overline{\widetilde{N}}_g)_u$  count the number of left--moving oscillators acting on the state. 
Only the states that survive the orbifold projection conditions, 
\equ{
P_{g} \cdot V_{g'}-p_{g} \cdot v_{g'} \equiv  
\frac{1}{2}\big(V_{g} \cdot V_{g'}-v_{g} \cdot v_{g'}\big) 
+ \big(\overline{\widetilde{N}}_{g}-\widetilde{N}_{g}\big) \cdot v_{g'} 
+\frac{\gve^{\!\times}}{2} \big(t_{1}^{\phantom{\prime}} t_{2}^{\prime}-t_{2}^{\phantom{\prime}} t_{1}^{\prime}\big)~, 
}
are part of the physical orbifold spectrum. 
The last term indicates that discrete torsion only affect the twisted sectors:  
the resulting twisted spectra are each others charge conjugates. 
Hence, discrete torsion leads to distinct spectra and hence interactions in the energy effective theory in four dimensions. 
This has, in particular, consequences for which blowup modes are available, see e.g.\cite{Faraggi:2021fdr}. 
For the purpose of this proceedings only the twisted states with oscillator excitations in the three twisted sectors with $t=(t_1,t_2): (1,0), (0,1)$ and $(1,1)$ are given in Table~\ref{tb:TwistedSpectrumWithOutOscillators}.

There exists a number of different $T^6/\Intr_2\times\Intr_2$ orbifolds depending on their underlying six--torus lattice, see {\em e.g.}\ refs.\cite{Faraggi:1992yz, Faraggi:2006bs, Donagi2008,FRTV,Athanasopoulos:2016aws}\,; here the orbifold with Hodge numbers $(51,3)$ is considered only. 
Therefore, each twisted sector of this orbifold has support at 16 fixed two--tori. 
The fixed two--tori of sector (1,0) are labelled by $y,z=1,\ldots, 4$, which determine their positions in the second and third two--tori directions. 
Similarly, the positions of the fixed two--tori of sector (0,1) are labelled by $x,z=1,\ldots, 4$  in the first and third two--tori directions, and  the positions of the fixed two--tori of sector (1,1) are labelled by $x,y=1,\ldots, 4$ in the first and second two--tori directions.

\begin{table}[h]
\tbl{\label{tb:TwistedSpectrumWithOutOscillators}
This table lists the shifted momenta $p_g, P_g$ of bosonic twisted states with left--moving oscillator excitations (complex conjugated oscillators are not made explicit here) and indicates whether they are in the physical spectrum without or with torsion, $\ge^{\!\times} =0$ or $1$, respectively. }
{
\begin{tabular}{@{}ccccc@{}}
\toprule
Sector & Bosonic twisted states $|p_g,P_g\rangle$ with oscillator excitations  & Repr. & $\gve^{\!\times} =0$ & $1$ 
\\ \colrule
 \multirow{ 2}{*}{$(1,0)$}  & 
$\tga_{-\sfrac12}^{2}
\Big|\big(0,\sfrac12,\sfrac12\big)\big(0,\sm\sfrac{1}{2},\sfrac{1}{2}, 0^{5}\big)\big(0^{8}\big)\Big\rangle; 
\tga_{-\sfrac12}^{3}
\Big|\big(0,\sfrac12,\sfrac12\big)\big(0,\sfrac{1}{2},\sm\sfrac{1}{2}, 0^{5}\big)\big(0^{8}\big)\Big\rangle$
& $4\,(1)$ & in & out 
\\ 
& 
$\tga_{-\sfrac12}^{2}
\Big|\big(0,\sfrac12,\sfrac12\big)\big(0,\sfrac{1}{2},\sm\sfrac{1}{2}, 0^{5}\big)\big(0^{8}\big)\Big\rangle; 
\tga_{-\sfrac12}^{3}
\Big|\big(0,\sfrac12,\sfrac12\big)\big(0,\sm\sfrac{1}{2},\sfrac{1}{2}, 0^{5}\big)\big(0^{8}\big)\Big\rangle$
& $4\,(1)$ & out & in   
\\ \colrule
\multirow{ 2}{*}{$(0,1)$}  
& 
$\tga_{-\sfrac12}^{1}
\Big|\big(\sfrac12,0,\sfrac12\big)\big(\sm\sfrac{1}{2},0,\sfrac{1}{2}, 0^{5}\big)\big(0^{8}\big)\Big\rangle; 
\tga_{-\sfrac12}^{3}
\Big|\big(\sfrac12,0,\sfrac12\big) \big(\sfrac{1}{2},0,\sm\sfrac{1}{2}, 0^{5}\big)\big(0^{8}\big)\Big\rangle$ 
& $4\,(1)$ & in & out
\\ 
& 
$\tga_{-\sfrac12}^{1}
\Big|\big(\sfrac12,0,\sfrac12\big)\big(\sfrac{1}{2},0,\sm\sfrac{1}{2}, 0^{5}\big)\big(0^{8}\big)\Big\rangle; 
\tga_{-\sfrac12}^{3}
\Big|\big(\sfrac12,0,\sfrac12\big)\big(\sm\sfrac{1}{2},0,\sfrac{1}{2}, 0^{5}\big)\big(0^{8}\big)\Big\rangle$
& $4\,(1)$ & out & in    
\\ \colrule
 \multirow{2}{*}{$(1,1)$}  
& 
$\tga_{-\sfrac12}^{1}
\Big|\big(\sfrac12,\sfrac12,0\big)\big(\sm\sfrac{1}{2},\sfrac{1}{2},0, 0^{5}\big)\big(0^{8}\big)\Big\rangle; 
\tga_{-\sfrac12}^{2}
\Big|\big(\sfrac12,\sfrac12,0\big)\big(\sfrac{1}{2},\sm\sfrac{1}{2},0, 0^{5}\big)\big(0^{8}\big)\Big\rangle$ 
& $4\,(1)$ & in & out 
\\
& 
$\tga_{-\sfrac12}^{1}
\Big|\big(\sfrac12,\sfrac12,0\big)\big(\sfrac{1}{2},\sm\sfrac{1}{2},0, 0^{5}\big)\big(0^{8}\big)\Big\rangle; 
\tga_{-\sfrac12}^{2}
\Big|\big(\sfrac12,\sfrac12,0\big)\big(\sm\sfrac{1}{2},\sfrac{1}{2},0, 0^{5}\big)\big(0^{8}\big)\Big\rangle$ 
& $4\,(1)$ & out & in 
\\ \botrule
\end{tabular}  
}
\end{table}

\section{GLSMs for standard embedding resolutions of $\boldsymbol{T^6/\Intr_2\times\Intr_2}$}
\label{sc:T6Z22glsms}

%
%
%
%
%

A GLSM description of the resolution of toroidal orbifold $T^6/\Intr_2\times\Intr_2$ in the standard embedding can be obtained as follows\cite{Blaszczyk:2011hs}\,: 
\begin{enumerate}
\item Start from a GLSM description of the underlying six--torus $T^6$ compatible with the $\Intr_2\times\Intr_2$ orbifold symmetries.
\item Add so--called exceptional gaugings to introduce the orbifold actions and define the exceptional cycles.
\end{enumerate}
This program is executed in a $(0,2)$ language, as the torsion models constructed later cannot be described using $(2,2)$ GLSMs.  
Therefore, first $(0,2)$ GLSMs encoding the geometry are determined. 
Then by insisting on implicit $(2,2)$ worldsheet supersymmetry, additional superfields, fermionic gaugings and superpotentials are inferred.

\subsection{$T^6$ GLSM with ${\Intr_2\times \Intr_2}$ symmetries}
\label{sc:TwoToriGLSM}

\begin{table}[h]
\tbl{\label{tb:T2wZ2glsm}
Gauge charges of the $(0,2)$ Superfields of a  $T^6$ GLSM with $\Intr_2\times \Intr_2$ symmetries. }
{
\begin{tabular}{@{}ccccccc@{}}
\toprule
\text{Sfield} & $\gF_{u\,1}$ & $\gF_{u\,2}$ & $\gF_{u\,3}$ & $\gF_{u\,4}$ & $\gG^{\phantom{\prime}}_u$ & $\gG_u' $ 
\\ 
 & $\gL_{u\,1}$ & $\gL_{u\,2}$ & $\gL_{u\,3}$ & $\gL_{u\,4}$ & $\gPs^{\phantom{\prime}}_u$ & $\gPs_u' $ 
\\ \colrule
$R_{u'}$ & $\sfrac 12\gd_{u'u}$ & $\sfrac 12\gd_{u'u}$ & $\sfrac 12\gd_{u'u}$ & $\sfrac 12\gd_{u'u}$ & $-\gd_{u'u}$ & $-\gd_{u'u}$ 
\\ \botrule
\end{tabular}
}
\end{table}

Following ref.~\cite{Blaszczyk:2011hs,Faraggi:2022gkt} toroidal orbifolds $T^6$ with $\Intr_2\times\Intr_2$ symmetries can be described by the superfields given in Table~\ref{tb:T2wZ2glsm}.  
The chiral superfields $\gF_{u\, x}$ and the fermi superfields $\gG_u, \gG^\prime_u$ encode the geometry. The fermi superfields $\gL_{u\, x}$ and the chiral superfields $\gPs_u, \gPs^\prime_u$ describe the gauge bundle in the $(0,2)$ language. 
By $(2,2)$ supersymmetry the charges of Table~\ref{tb:T2wZ2glsm} imply the fermionic gauge symmetries 
\equ{ \label{eq:FermiGaugeBundleSimple} 
\gd \gL_{u\,x} = \sfrac 12\, \gF_{u\,x} \,\gX_u~, 
\qquad  
\gd \gG_u = - \gPs_u\, \gX_u~, 
\qquad 
\gd \gG_u' = - \gPs_u'\, \gX_u~, 
}
where $\gX_u$ are the fermi superfield gauge parameters\footnote{The appearance of fermionic gauge symmetries can be understood as a consequence of decomposing (2,2) super gauge transformations into the (0,2) language: a (2,2) chiral super gauge parameter splits up into a scalar and a fermi chiral (0,2) superfield\cite{Witten:1993yc}\,.}. 
The elliptic curve descriptions of the three two tori are cut out by the F--term constraints resulting from the superpotential 
\equ{ \label{eq:TwoTorusSuperpotential} 
P_\text{geom $T^6$} = \sum_u 
\Big( \gk_u\, \gF_{u\,1}^2 + \gF_{u\,2}^2 + \gF_{u\,3}^2\Big) \gG^{\phantom{\prime}}_u + 
\Big( \gF_{u\,1}^2 + \gF_{u\,2}^2 + \gF_{u\,4}^2\Big) \gG_u'~, 
}
where the complex structures $\gt_u$ of the three two--tori are parameterised 
\equ{ \label{eq:ComplexStructureTwoTori} 
\gk_u = 
\frac{\wp_{\gt_u}(\sfrac{\gt_u}{2}) - \wp_{\gt_u}(\sfrac 12)}
{\wp_{\gt_u}(\sfrac{1+\gt_u}{2}) - \wp_{\gt_u}(\sfrac 12)} 
}
 in terms of the Weierstrass $\wp$ function. 
(Details of this GLSM construction and its geometrical interpretation as a two--tori with $\Intr_2\times\Intr_2$ are discussed in ref.\cite{Blaszczyk:2011hs} in the (2,2) language.)
By $(2,2)$ supersymmetry there is an accompanying superpotential 
\equa{ \label{eq:TwoTorusSuperpotentialGauge} 
P_\text{bundle $T^6$} = &~ 2\sum_u 
\Big( \gk_u\, \gF_{u\,1} \gL_{u\,12}  + \gF_{u\,2} \gL_{u\,2} + \gF_{u\,3} \gL_{u\,3} \Big) \gPs^{\phantom{\prime}}_u  \non \\ 
+ &~ 2\sum_u  
\Big( \gF_{u\,1} \gL_{u\,1}  + \gF_{u\,2} \gL_{u\,2}  + \gF_{u\,4} \gL_{u\,3} \Big) \gPs_u'~. 
}
This expression is obtained by replacing the superfields $\gG_u, \gG_u'$ by $\gPs_u, \gPs_u'$ and applying the functional operator 
\equ{ \label{eq:FuncOperatorSimple} 
\sum_{u,x} \gL_{u\,x} \dfrac{\gd\phantom{\gF_{u\, x}}}{\gd\gF_{u\, x}}~. 
}

\subsection{Maximal full resolution GLSM for $T^6/\Intr_2\times\Intr_2$} 

There exist a whole family of resolution GLSMs that can be associated to a given orbifold with certain Hodge numbers\cite{Blaszczyk:2011hs}\,. 
In most of these resolution GLSMs not all exceptional K\"ahler parameters associated with the volumes of the exceptional cycles are made explicit. 
Physically this means either that the sizes of a number of exceptional cycles are identical or that not all exceptional cycles can be resolved.
The GLSM with the least number of gaugings which still resolves all singularities is called the minimal full resolution GLSM.
Instead, the maximal full resolution GLSM makes all K\"ahler parameters associated to the volumes of the ordinary and exceptional cycles are made explicit as FI parameters of the gaugings in the GLSM\footnote{Even though the maximal full resolution GLSM may be considered as the most complete GLSM description associated with a given orbifold, it might often be rather cumbersome to work with; the minimal full resolution GLSM is often more practical.}\,.

The maximal full resolution GLSM of the orbifold $T^6/\Intr_2\times\Intr_2$ with Hodge numbers $(51,3)$ has three ordinary gaugings $R_1$, $R_2$ and $R_3$ to define three two--tori and $3\cdot 16 = 48$ exceptional gaugings $E_{1,yz}$, $E_{2,xz}$ and $E_{3,xy}$ associated to the exceptional cycles. 
This GLSM should therefore be able to continuously interpolate between all $4^{64}$ fully smooth resolutions of the $T^6/\Intr_2\times\Intr_2$, distinguished by their triple intersection numbers, by varying these $51$ K\"ahler parameters.
The charge table of the maximal full resolution GLSM is given in Table~\ref{tb:MaxT6Z22glsm}.

The minimal full resolution GLSM can be easily inferred from the maximal version: one simply drops all labels $x,y,z$ that distinguish the different fixed tori and exceptional divisors. Hence, in particular, the minimal full resolution GLSM has one three exceptional gaugings $E_1, E_2, E_3$: the K\"ahler parameters of 16 exceptional divisors are all taken to be equal.

\begin{table}[h]
\tbl{\label{tb:MaxT6Z22glsm} 
Gauge charges of the maximal full resolution GLSM for $T^6/\Intr_2\times\Intr_2$. 
(The normalization of the gauge charges of the exceptional gauging is chosen to follow the values of the shifted momenta of the twisted states and this normalization is extended to the ordinary gaugings $R_1,R_2, R_3$ as well.)}
{
\begin{tabular}{@{}c ccc c c c ccc@{}}
\toprule
\text{Sfield} & $\gF_{1\,x}$ & $\gF_{2\,y}$ & $\gF_{3\,z}$ & $\gG_1, \gG_1'$ & $\gG_2,\gG_2'$ & $\gG_3, \gG_3'$ & $\gF_{1\,yz}'$ & $\gF_{2\,xz}'$ & $\gF_{3\,xy}'$
\\ 
 & $\gL_{1\,x}$ & $\gL_{2\,y}$ & $\gL_{3\,z}$ & $\gPs_1,\gPs_1'$ & $\gPs_2,\gPs_2'$ & $\gPs_3,\gPs_3'$ & $\gL_{1\,yz}'$ & $\gL_{2\,xz}'$ & $\gL_{3\,xy}'$  
\\ \colrule
$R_1$ & $\sfrac 12$ & $0$ & $0$ & $-1$  & $0$  & $0$ & $0$ & $0$ &$ 0$ 
\\
$R_2$ & $0$ & $\sfrac 12$ & $0$ & $0$ & $-1$ & $0$  & $0$ & $0$ & $0$ 
\\ 
$R_3$ & $0$ & $0$ & $\sfrac 12$ & $0$ & $0$ & $-1$  & $0$ & $0$ & $0$ 
\\ \colrule
$E_{1\,y'z'}$ & 0 & $\sfrac 12 \gd_{y'y}$ & $\sfrac 12 \gd_{z'z}$  & $0$ & $0$ & $0$ & $-\gd_{y'y}\gd_{z'z}$ & $0$ & $0$ 
\\ 
$E_{2\,x'z'}$ & $\sfrac 12\gd_{x'x}$ & $0$ & $\sfrac 12 \gd_{z'z}$  & $0$ & $0$ & $0$ & $0$ & $-\gd_{x'x}\gd_{z'z}$ & $0$ 
\\  
$E_{3\,x'y'}$ & $\sfrac 12 \gd_{z'z}$ & $\sfrac 12 \gd_{y'y}$ & $0$ & $0$ & $0$ & $0$ & $0$ & $0$ & $-\gd_{x'x}\gd_{y'y}$ 
\\ \botrule
\end{tabular} 
}
\end{table}

This charge table is in line with the description of local orbifold resolutions using twisted states with oscillators as proposed in ref.\cite{GrootNibbelink:2010qut}\,: 
the shifted momenta $p_g$, given in Table~\ref{tb:TwistedSpectrumWithOutOscillators} for the various twisted sectors, determine the exceptional charges $E_{r\,xy}$ of the chiral multiplets $\gF_{u\,x}$ completed with novel chiral multiplets $\gF_{r\,xy}'$ with charge $-1$ for each of the three twisted sectors $r$ located at the fixed two--tori labeled by $x,y$.
The GLSM charges of the fermi multiplets cannot directly be determined by the shifted momenta $P_g$ given in Table~\ref{tb:TwistedSpectrumWithOutOscillators}. 
To obtain the charges of the fermi multiplets $\gL_{u\,x}$ one has to add $+1$ to the negative entry of $P_g$ and add novel fermi multiplets $\gL_{r\, xy}'$ with charge $-1$.
The choice of these signs is such that the charges of the fermi multiplets $\gL_{u\,x}$ and $\gL_{r\,xy}'$ match those of the chiral multiplets $\gF_{u\,x}$ and $\gF_{r\,xy}'$ in order to comply with $(2,2)$ supersymmetry. 
In order that the number of fermionic degrees of freedom remain the same, the exceptional gaugings have to be accompanied by fermionic gauge transformations. 
Given the charge assignment in Table~\ref{tb:MaxT6Z22glsm} the following specific form of the fermionic gaugings 
\equ{
\arry{ccc}{
\gd \gL_{1\,x} = \sfrac 12\, \gF_{1\,x} \Big(\gX_1 +\sum\limits_z \gY_{2\,xz} + \sum\limits_y \gY_{3\,xy}\Big)~, 
&\qquad& 
\gd \gL_{1\,yz}' = - \gF_{1\,yz}'\, \gY_{1\,yz}~,
\\
\gd \gL_{2\,y} = \sfrac 12\, \gF_{2\,y} \Big(\gX_2 + \sum\limits_z \gY_{1\,yz} + \sum\limits_x \gY_{3\,xy}\Big)~, 
&&
\gd \gL_{2\,xz}' = - \gF_{2\,xz}'\, \gY_{2\,xz}~, 
\\
\gd \gL_{3\,z} = \sfrac 12\, \gF_{3\,z} \Big(\gX_3 + \sum\limits_y \gY_{1\,yz} + \sum\limits_x \gY_{2\,xz}\Big)~, 
&& 
\gd \gL_{3\,xy}' = - \gF_{3\,xy}'\, \gY_{3\,xy}~, 
\\
\gd\gG_u = - \gPs_u\, \gX_u~, 
\qquad 
\gd\gG_u' = - \gPs_u'\, \gX_u~, 
&& 
}
 \label{eq:FermiGaugeBundle} 
}
where $\gY_{r\, xy}$ denote the fermi gauge parameters associated with the exceptional gaugings $E_{r\,xy}$ by $(2,2)$ supersymmetry. In addition, there are thirteen of neutral fermi multiplet $\gL^0_p$, $p=1,\ldots, 13$, to complete the heterotic theory, corresponding to the $5+8=13$ zero entries in the shifted momenta $P_g$ given in Table~\ref{tb:TwistedSpectrumWithOutOscillators}.

The fermi superfields $\gG_1, \gG_1'$, $\gG_2, \gG_2'$ and $\gG_3,\gG_3'$ feature in the superpotential to define the three underlying two--tori, see~\eqref{eq:TwoTorusSuperpotential}. 
In light of the exceptional gaugings the superpotential is extended to 
\equ{
P_\text{geom\ res}   = \frac 12
\sum\limits_{r\neq s\neq u} \prod\limits_{y,z} 
\Big\{ 
\big( 
\!\gk_u\, \gF_{u\,1}^2  \gF_{r\,1z}' \gF_{s\,1y}' 
+ \gF_{u\,2}^2  \gF_{r\,2z}'  \gF_{s\,2y}' 
+ \gF_{u\,3}^2 \gF_{r\,3z}' \gF_{s\,3y}'
\big) \gG_u
\non \\[1ex] 
+ 
\big( 
\gF_{u\,1}^2 \ \gF_{r\,1z}'  \gF_{s\,1y}' 
+ \gF_{u\,2}^2  \gF_{r\,2z}' \gF_{s\,2y}' 
+ \gF_{u\,4}^2 \gF_{r\,4z}' \gF_{s\,4y}' 
\big) \gG_u'\Big\}
 \label{eq:T6Z22maxSuperpotential} 
} 
together with a bundle superpotential $P_\text{bundle\ res}$. 
It can be obtained from the geometry superpotential in a similar fashion as 
\eqref{eq:TwoTorusSuperpotentialGauge} was obtained by replacing and superfields $\gG_u, \gG_u'$ by $\gPs_u, \gPs_u'$ and applying the functional operator 
\equ{
\sum_{u,x} \gL_{u\,x} \dfrac{\gd\phantom{\gF_{u\, x}}}{\gd\gF_{u\, x}} + 
\sum_{r,x,y} \gL_{r\,xy}' \dfrac{\gd\phantom{\gF_{r\, xy}'}}{\gd\gF_{r\, xy}'}~, 
}
which extends  the operator~\eqref{eq:FuncOperatorSimple}. 
Given that~\eqref{eq:T6Z22maxSuperpotential} is already a bulky expression, the resulting bundle superpotential is not explicitly given here. 

Notice that the resulting GLSM is very similar as the maximal full resolution GLSM discussed in ref.~\cite{Faraggi:2022gkt}\,. 
However, there the local resolutions could be thought of as line bundle backgrounds since there were no fermionic gaugings associated to the exceptional gaugings and no novel fermi multiplets $\gL_{r\,xy}'$ were added.

\section{GLSMs resolutions of $\boldsymbol{T^6/\Intr_2\times\Intr_2}$ with torsion}
\label{sc:T6Z22torsion}

\subsection{Construction of GLSM associated to the torsion orbifold}

The effect of discrete torsion is that the opposite charged twisted states survive the orbifold projections.  This implies that the exceptional gauge charges $E_{r\,xy}$ of the fermi multiplets $\gL_{u\,x}^{\!\!\times}$ and $\gL_{r\,xy}^{\!\!\times\,\prime}$ in the GLSM associated to the orbifold with torsion are precisely opposite to those of $\gL_{u\,x}$ and $\gL_{r\,xy}'$  in the GLSM associated to the orbifold model without torsion\footnote{The superscript $\times$ distinguishes the fermi superfields obtained from the orbifold with torsion from the fermi superfields obtained from the orbifold without torsion discussed in the previous section.}, while all other superfields (in particular $\gF_{u\,x}$ and $\gF_{r\,xy}'$) and all other charges (in particular the $R_u$ charges of  $\gL_{u\,x}$ and $\gL_{r\,xy}'$) are kept the same.

Hence, the torsion resolution GLSM is a genuine $(0,2)$ model that cannot be lifted to a $(2,2)$ theory: for example the fermi multiplets $\gL_{u\,x}^{\!\!\times}$ has the opposite exceptional charges to the chiral multiplets $\gF_{u\,x}$, hence they cannot form $(2,2)$ multiplets anymore. 
To obtain the expressions for the fermionic gauge transformations and the superpotentials, the following replacements 
\equ{ \label{eq:ReplacementsTorsion} 
\gL_{u\,x} \ra \prod_y \gF_{r\,xy}^{\prime -1} \prod_z \gF_{s\,xz}^{\prime -1} \gL_{u\,x}^{\!\!\times}~, 
\qquad
\gL_{r\,xy}^\prime \ra \gF_{r\,xy}^{\prime 2} \gL_{r\,xy}^{\!\!\times\,\prime}~, 
}
with $r\neq s\neq u$, should be applied to the appropriate expressions in the non--torsion theory. 
The powers  of $\gF_{r\,xy}^{\prime}$ are dictated by the difference in exceptional gauge charges of the fermi multiplets in the models associated to the orbifolds with and without torsion.
Applying this replacement to the fermionic transformation rules~\eqref{eq:FermiGaugeBundle} leads, in particular, to the transformations 
\equ{ 
\gd \gL_{r\,xy}^{\!\!\times\,\prime} = - \gF_{r\,xy}^{\prime\,-1}\, \gX_{r\,xy}~,  
}
which are not holomorphic at the loci $\gF_{r\,xy}'=0$. However, as discussed next, these loci are avoided because of logarithmic singularities that arise at them due to field dependent FI--terms.

\subsection{Cancellation of mixed anomalies}

The sign--flipped $E_{r\,xy}$--gauge charges do not affect the anomaly cancellation conditions for the $R_u, R_{u'}$ gaugings as well as the expectional gaugings $E_{r\,xy}, E_{r'\,x'y'}$ as these pure and mixed anomalies are even under these sign-flips. 
However, the mixed $R_u E_{r\neq u\,xy}$--anomalies do not vanish
\equ{
\cA_{u~r\,xy} = \cA_{r\,xy~u} = \sfrac 12\cdot \sfrac 12 - (\sfrac12)\cdot (-\sfrac 12) = \sfrac 12~, 
} 
$u \neq r$. 
This can be easily confirmed using Table~\ref{tb:MaxT6Z22glsm} taking into account that the $E_{r\,xy}$ charges of fermi multiplets $\gL_{u\,x}^{\!\!\times}$ and $\gL_{r\,xy}^{\!\!\times}$ have opposite signs. 
Hence, the GLSMs associated to the $T^6/\Intr_2\times\Intr_2$ orbifold models without or with torsion are genuinely physically distinct. 

In terms of the super field strengths $F_u$ and $F_{r\,xy}$ of the $R_u$-- and $E_{r\,xy}$--gaugings, the field dependent FI--terms~\cite{Blaszczyk:2011ib,Quigley:2011pv}
\equ{ \label{eq:FIanom} 
W_\text{FI\,anom} = \frac 1{4\pi} \sum_{u\neq r,x,y} 
\Big( 
\frac{1-c_{u~r\,xy}}2\, \log (N_{r\,xy})\, F_u + 
\frac{c_{u~r\,xy}}2 \,\log (N_u)\, F_{r\,xy}
\Big)~, 
}
cancel these mixed anomalies provided that  the chiral superfield funtions $N_{r\,xy}$ and $N_u$ have negative unit charge under the $E_{r\,xy}$-- and $R_u$--gaugings, respectively, and all other gauge charges are  zero. 
Since by counter terms can shift the two--dimensional mixed anomalies around\footnote{These counter terms are supersymmetrizations of gauge field one--form bi--linears as discussed around equation (17) in ref.\cite{Blaszczyk:2011ib}\,.}, these field dependent FI--terms contain arbitrary coefficients $c_{u~r\,xy}$. 
The explicit expressions of these functions $N_{r\,xy}$ and $N_u$ involve various parameters that encode the positions of the NS5--branes~\cite{Blaszczyk:2011ib,Quigley:2011pv}\,. 
A particularly simple representation is obtained by taking all $c_{u~r\,xy}=0$, so that the functions $N_u$ become irrelevant, and set 
\equ{
N_{r\,xy} = \gF_{r\,xy}'~. 
}
For this choice the NS5--branes are all located at the resolved exceptional divisors $E_{r\,xy}$.

\section{Discusions} 
\label{sc:Discusions}

This proceedings studied the effect of discrete torsion between the two orbifold twists of the $T^6/\Intr_2\times\Intr_2$ orbifold in blowup using a $(0,2)$ GLSM formalism. 
To this end first the twisted states that generate the resolution were identified in the orbifold theory. 
These states with oscillator excitations were chosen which, in blowup, are described by GLSMs with exceptional gaugings accompanied by fermionic gauge symmetries. 
Consequently, the model without torsion effectively possesses $(2,2)$ supersymmetry. 

Since the effect of the discrete torsion is that opposite charged twisted states survive the orbifold projections, the fermi multiplets in the resolution GLSMs without and with torsion have opposite exceptional gauge charges. 
Even though this might seem insignificant, it results in a number of mixed anomalies between the exceptional gaugings and the gaugings that are used in the definition of the elliptic curve descriptions of the underlying $T^6$. 
These mixed anomalies can be cancelled by field dependent FI--terms implementing a worldsheet version of the Green--Schwarz mechanism. 
Since these FI--terms become logarithmically divergent if some (combination of) chiral superfields vanish and hence signify that the heterotic worldsheet is interacting with NS5--branes. 
For a specific choice of parameters encoding different realisations of these field dependent FI--terms, these NS5--branes are located at the exceptional cycles that appeared in the resolution process. 

In order that a CFT interpretation of these GLSMs are viable, it is necessary that global left- and right-symmetries can be defined. Since the sums of charges of each gauge symmetry of all chiral or fermi superfields separately vanish, this is guaranteed. This happens independently on whether the GLSM is associated to the torsional orbifold or not, since the effect of the discrete torsion was the flip of all exceptional gauge charges of all fermi multiplets, hence such sums vanish irrespectively of whether torsion is switched on or not. 

This proceedings is closely related to the publication ref.~\cite{Faraggi:2022gkt} by the author. 
In that work twisted states without oscillator excitations were considered to generate the resolutions, while in this work twisted states with oscillators were used. 
As a consequence, in that work only the minimal full resolution GLSM could be considered as otherwise one would need more charged fermi superfields than the theory possesses. 
This problem did not occur in this work as GLSMs, generated by twisted states with oscillators, require fermionic gaugings accompanying the exceptional gaugings of the local blowups. 
Therefore, the aforementioned problem did not arise in the present work for the maximal full resolution GLSM. 
As far as the distinction between the GLSMs associated with the orbifold without and with torsion goes, the findings in this proceedings and ref.~\cite{Faraggi:2022gkt} essentially coincide: 
the choice of twisted states without or with oscillator excitation did not seem to affect the torsion resolution GLSMs in a significant way. 
In both cases mixed anomalies arise which can be cancelled by very similar field dependent FI--interactions indicating that NS5--branes are present.

\section*{Acknowledgments}

The author would like to thank the organisers of the workshop GLSM@30 and the Simons Center for their kind invitation and for hosting this very pleasant workshop and the opportunity to contribute to the proceedings of this meeting. 
In addition, the author is indebted to Alon Faraggi and Martin Hurtado Heredia for the inspiring collaboration on which a large part of this work has been based. 
Finally, the author would like to thank Benjamin Percival for carefully reading the manuscript and providing valuable comments.

\bibliographystyle{ws-ijmpa}
 \bibliography{proceedings}

\end{document}